\documentstyle[12pt,aps,prd,graphicx]{revtex}

\begin{document}

\title{ Formation and   Evaporation  of Charged Black Holes}
\author{Evgeny Sorkin$^{\ast}$ and Tsvi Piran$^{\dagger}$}
\address{The Racah Institute of Physics, The Hebrew University,
  Jerusalem, Israel, 91904}
 \date{26 February 2001}
\maketitle 
\begin{abstract} 
  We investigate the dynamical formation and evaporation of a
  spherically symmetric charged black hole. We study the
  self-consistent one loop order semiclassical back-reaction problem.
  To this end the mass-evaporation is modeled by an expectation value
  of the stress-energy tensor of a neutral massless scalar field,
  while the charge is not radiated away. We observe the formation of
  an initially non extremal black hole which tends toward the extremal
  black hole $M=Q$, emitting Hawking radiation.  If also the discharge
  due to the instability of vacuum to pair creation in strong electric
  fields occurs, then the black hole discharges and evaporates
  simultaneously and decays regularly until the scale where the
  semiclassical approximation breaks down. We calculate  the
  rates of the mass and the charge loss and estimate the life-time of 
  the decaying black holes.

\end{abstract}

\section{Introduction}
\label{intro}

Extremal black holes (EBHs) play an important role in black hole
thermodynamics. The vanishing of the surface gravity of such black
holes implies zero Hawking-Bekenstein temperature. This suggests on
the tight connection of EBH in black hole physics with zero
temperature states in thermodynamics. The zero temperature EBH is,
however, unattainable in a finite sequence of physical processes
\cite{isr}.  This means that the dynamical evolution of a non extremal
black hole towards the EBH state continues an infinite (advanced)
time.  The full non linear evolution towards an EBH presents a  step
towards understanding the status of EBHs in the black hole
thermodynamics. The solution provides the entire dynamical spacetime
which emerges in this case.

A generic black hole  carries an angular momentum and an electric
charge. The formation and evaporation of such a black hole is a
challenging and extremely difficult problem both analytically and
numerically. That is because the problem  is necessary  three
dimensional. Therefore, we consider here a simpler spherically
symmetric problem of a formation and a decay of a charged black hole.
The EBHs in this simple model have all the features of a more general
Kerr-Newman black hole.  In earlier studies \cite{ted,StTri,LooL} of
the evaporation of Reissner-Nordstr\"{o}m (RN) black holes an
initially singular spacetime--the static extremal RN black hole--was
considered.  This extremal black hole was exited above the extremality
by an infalling matter and then allowed to decay back toward the $Q=M$
state by emitting Hawking radiation.  Here we pose the problem of the
evaporation of a RN black hole in somewhat different fashion. We
formulate the problem of the formation and the evaporation of a
spherically symmetric charged black hole beggining with the 
collapse of an initially regular self-gravitating charged matter
distribution.  We include quantum effects, which are considered here
in the one loop order, in a self-consistent manner.  To do so, we
introduce the expectation value of a scalar field stress-energy
tensor, which acts as an additional source in  the semiclassical
Einstein equations (see e.g. \cite{BirDev}).  We regard the charge as
 stable, not allowing it to be radiated away \footnote{This can
  be realized in practice when the charged particles are sufficiently
  massive and there is no significant creation of them in the Hawking
  or Schwinger processes.  }. The semiclassical equations are
integrated numerically to give the structure of the spacetime for this
configuration. We find that the evaporation proceed to a stable
end-point corresponding to the extremal, $M=Q$, charged black hole.

In the next stage we relax the assumption
of a stable charge, allowing the black hole to discharge via the
Schwinger pair creation process in addition to Hawking evaporation.
The characteristic quantity governing the Schwinger process is the
critical pair-creating field, $E_c \equiv { \pi m^2 c^3 / e
  \hbar}$, with $e$ and $m$ are the charge and the mass of the
produced particles. Suppose that the critical field is reached
below the maximal radius of the outer apparent horizon (i.e. the
apparent radius of a non-evaporating black hole), but before the EBH
state is approached.  In this arrangement the black hole evaporates,
radiating away its mass, the outer apparent horizon shrinks until the
electric field upon this horizon reaches the critical pair-creating
value. Then  electrically charged pairs are produced intensively. 
Particles having the same polarity as a black hole are repulsed to
infinity. Particles with the opposite  polarity are captured by the
black hole reducing its charge. Therefore, instead of
reaching the stable EBH state the black hole evaporates and discharges
simultaneously.  The black hole looses its mass and the
charge, disappearing completely (until it reaches scales where the the
semiclassical approximation breaks down). It is of interest to note
that the life-time of such decaying black hole is longer than that of
a neutral  evaporating black hole of the same mass.

The article is organized as follows.  In section \ref{model} we
introduce the physical model of and write down the evolutionary
equations.  The characteristic problem and the numerical scheme are
described in section \ref{numerics}.  Section \ref{results} is devoted
to our numerical and analytical results and in section
\ref{conclusions} we summarize our findings.  We use units in which
$c=G=\hbar=1$.

\section{The Model}
\label{model}
We introduce a simple spherically-symmetric four dimensional model
which captures the essence of the realistic behavior of the system.
The semiclassical back-reaction is included in one loop order by
introducing the expectation value of the stress-energy tensor, which
arises from the quantization of a $2D$ massless scalar field in $2D$
curved spacetime \cite{DFU}.  To mimic the $4D$ radial dependence we
divide this tensor by $4\pi r^2$.  This stress-energy tensor offers a
good approximation to the full four dimensional theory since (i) most
of the energy is carried away in $S-$waves \cite{Page}, and (ii) the inclusion
of higher angular modes does not change significantly the behavior
derived from the $S-$wave approximation \cite{ted}, see, however
\cite{StTri}.  Even though the tensor, which we utilize, arises in a
$2D$ theory and not from the spherical reduction \cite{BalFa} of the
full $4D$ theory, it gives a feeling of what the back-reaction would
be, and, indeed, this tensor engenders the evaporation of black holes
(see e.g. Ref.  \cite{ParPir,AyPir}).

Our final aim is the numerical integration of the evolution equations.
To simplify the numerical procedure we apply several convenient tools.
The charged collapsing matter is simulated by a complex massless
scalar field. The ``massless'' nature of the fields makes it very
convenient to use the double-null coordinates.  The line element is
 taken to be of the form:
\begin{equation}
\label{metric}
{{ds}^2=-{{\alpha(u,v)}^2 du dv} +{r(u,v)}^2 d{\Omega}^2     }  \ ,
\end{equation}
where $d{\Omega}^2$ is the unit two-sphere.  There is a coordinates
gauge freedom. For the time being  ${u,v}$ are just general light-cone
coordinates which would be specified by fixing a gauge (see section
\ref{numerics}).

We formulate the set of coupled Einstein-Maxwell-Klein-Gordon
equations as a first order system of PDEs. The numerical integration of the
first order system functions very well both for the neutral
\cite{HamSt}, and for the charged \cite{HodPir,SorPir} situations.  It is
convenient to define the auxiliary variables:
\begin{equation}
\label{newvar}
d \equiv{\alpha_v\over\alpha}, \ h\equiv{\alpha_u\over\alpha}, \
f\equiv r_u,  \   g\equiv r_v, \ 
  w\equiv s_u,\     z\equiv s_v \ ,
\end{equation} 
wherein $s$ is the complex scalar field divided by $\sqrt{4\pi}$.  We
have adopted the notation $W_x \equiv {\partial W / \partial x} $ for
partial derivatives of any function $W=W(x,y)$.

The Hawking radiation is modeled by the expectation value of the
quantum stress-energy tensor of a $2D$ massless scalar field.  In the
light-cone coordinates (\ref{metric}) the only non vanishing
components of this tensor (divided by $4\pi r^2$)  are:
\begin{eqnarray}
\label{tensor}
\langle T_{uv} \rangle &=& \langle T_{vu} \rangle =-{P \over r^2}  d_u, \\
\langle T_{uu} \rangle &=&{P \over r^2}(h_u-h^2), \\
\langle T_{vv} \rangle &=&{P \over r^2}(d_v-d^2) \ ,
\end{eqnarray} 
wherein $P$ is a constant, which is proportional to the number of
massless scalar fields. $P$ controls the rate of the evaporation and
it also defines the length scale where the semiclassical
approximation breaks down.

We write the coupled set of the semiclassical Einstein-Maxwell-KG equations:
\noindent
                   
Einstein equations:
\begin{eqnarray*}
\label{einstein}
 E1  &:&     d_u   =  {f g \over r^2 } +{\alpha^2 \over 4 r^2}-
                          {\alpha^2q^2 \over 2 r^4}- 
                         {1\over 2}(w z^*+w^* z )-{1\over 2}i e a(s
                         z^* - s^* z)  \ ,\\
E2  &:&     g_v  =  2 d g  -  r z^*z   -
                         { P\over r}(d_v-d^2) \ ,\\ 
E3  &:&      f_v  =  - {f g \over r}   -  {1\over 4 r}\alpha^2   +
                  {\alpha^2 q^2 \over 4 r^3}- \\
   &-& {P \over4 r}\left[ {f g \over r^2 } +{\alpha^2 \over 4 r^2}-
                          {\alpha^2q^2 \over 2 r^4}    
                         -{1\over 2}(w z^*+w^* z )-{1\over 2}i e a(s
                         z^* - s^* z)\right]  \ .
\end{eqnarray*}
Maxwell equations:
\begin{eqnarray*}
M1 &\equiv& a_v - {\alpha^2q \over 2 r^2} = 0 \ , \\
M2  &\equiv&  q_v - i e r^2(s^* z -s z^*) =0  \ .
\end{eqnarray*}
Where $a(u,v)=A_u$ is the  electromagnetic potential  and $q(u,v)$ is the
charge.  The complex scalar field (Klein-Gordon) equations expand to:
\begin{eqnarray*}
S1 &\equiv&   r z_u  + f z  + g w  +  i e a r z  + i e a g s +
                    {i e \over 4 r}\alpha^2 q s = 0 \ , \\
S2  &\equiv& r w_v  +  g w  + f z  +  i e a r z  + i e a g s +
                    {i e \over 4 r}\alpha^2 q s = 0 \ .
\end{eqnarray*}
Finally, the definitions (\ref{newvar}) are rewritten as:
\begin{eqnarray*}
D1 &\equiv& d-{\alpha_v\over\alpha}=0 \ ,  \\ 
D2 &\equiv& g-r_v=0 \  ,\\  
D3 &\equiv& z-s_v=0 \   .\\
\end{eqnarray*}
These equations should be accompanied by a specification of the
initial data and the suitable boundary conditions.
\section{Initial Setup and the Numerical Scheme}
\label{numerics}
The characteristic initial value problem for the complex scalar field
is a straightforward generalization of that given in Ref. \cite{BurOr}.
We have formulated it in  \cite{SorPir} and describe it briefly here.

We choose the initial characteristic surfaces to be: the ingoing
$v=const\equiv v_i $ hypersurface, and, the outgoing $u=const\equiv
u_i $ hypersurface.  The coordinate gauge freedom is fixed by
constraining the $r$ to be linear with $u$ or $v $ along the
characteristic hypersurfaces. Namely, on $u=u_i $ segment we choose
$g\equiv r_v=1$, on $v=v_i $ segment we choose $f\equiv r_u=r_{u0}$.
To get $r$ along initial surfaces it is necessary to supply
$r_0=r(u_i,v_i)$ that serves as a free parameter. For convenience we
choose $u_i=0$, $v_i=r_0$. Therefore, one obtains: $r(u_i,v)=v \ , \ 
r(u,v_i)=u r_{u0} + r_0$ .

The domain of integration in the $v$ direction confined in the compact
region: $ v_i \le v \le v_f $. The origin $r=0$, for simplicity, is
not included in the integration domain. This is achieved by an
appropriate choice of the final outgoing hypersurface $u_f$.  The
scalar field distribution is chosen to be non-vanishing only in the
compact segment $v_i \le v \le v_p$ ($v_p=v_2, v_2'$) and is taken in
a shape of a pulse, which is smoothly matched at the endpoints ($v_i$
and $v_p=v_2, v_2'$) to the rest of the integration domain:
\begin{equation}
s(u_i,v) =A \sin^2\left(\pi{v-v1 \over v_2-v1}\right) 
+i B  \sin^2\left(\pi{v-v1 \over v_2'-v1}\right) \ ,
\end{equation}
where $A,B$ are the amplitudes of the pulse. The initial values of
integrated variables are:
 \begin{eqnarray}
\label{z}
z(u_i,v)&=&   {A\pi\over v_2'-v_1} \sin\left(2\pi {v-v_1 \over v_2'-v_1}\right)
        +     {i B\pi\over v_2-v_1} \sin\left(2\pi {v-v_1\over v_2-v_1}\right) 
            \ ,\nonumber\\
w(u,v_i) & \equiv &0 \ .
\end{eqnarray}

The initial value for $d(u_i,v)$ can be obtained by a numerical
integration of the constraint equation $E2$ along the initial $u=u_i$
hypersurface, together with the choice $\alpha(u_i,v_i)=1$.  We
neglect the quantum stress-energy tensor on the initial hypersurfaces.
This is justified as we verify later in the numerical solution.  From
$E2$ we obtains the initial value for $d$:
\begin{eqnarray}
\label{d}
\lefteqn{d(u_i,v)=    {A^2\pi^2 v\over2 (v_2'-v_1)^2}
               \sin^2\left(2\pi {v-v_1 \over v_2'-v_1}\right) +} \nonumber\\
      & &  +   { B^2\pi^2 v\over2 (v_2-v_1)^2}
             \sin^2\left(2\pi {v-v_1\over v_2-v_1}\right)
\ .
\end{eqnarray}

With no sources the spacetime is  Minkowski spacetime for $v \le v_i $. The
boundary value for $\alpha(u,v_i)$ is obtained from the second
constraint equation $E3$ to be: $\alpha(u,v_i)=1$. The rest of the
boundary setup is $q(u,v_i)=0, a(u,v_i)=0$.

In our coordinates the mass-function is: $M(u,v)={r\over
  2}\left(1+{q^2\over r^2}+{4\over\alpha^2}r_u r_v\right)$. Since it
vanishes for Minkowski spacetime (in the region $v\le v_i$) one can
calculate: $r_{u0}=-{1\over 4 }$. It should be noted that for $v \gg
M$ our ingoing null coordinate $v$ is closely related (proportional)
to the ingoing Eddington-Finkelstein null coordinate $v_{\rm e}$. The
$u$ coordinate measures the proper time of an observer at the origin
\cite{SorPir}.

The numerical scheme used to evolve the initial data is similar to the
one used for the classical equations \cite{SorPir}.  At each step we
evolve $d$ and $z$ using $E1$ and $S1$ from the hypersurface $u$ to
$u+du$. We calculate the derivative $d_v$.  Then we solve the
appropriate equations for the rest of quantities along the outgoing
null rays $u+du=const$, starting from the initial outgoing
hypersurface $v=v_i$.  We integrate equation $D1$ to find $\alpha$,
then we solve the coupled differential equations $D2$ and $E2$ to
obtain $r$ and $g$. Next, equations $D3, M2, M1$ are integrated to
obtain $s, q$ and $a$.  Finally, the differential equations $E3$ and
$S2$ are solved for $f$ and $w$, respectively.

This integration scheme uses several different numerical methods
\cite{NumRec} to evolve the initial data.  To evolve the quantities in
the $u$-direction we utilize the 5-th order Cash-Karp Runge-Kutta
method.  The differential equations in the $v$ direction are solved
using a 4-th order Runge-Kutta method.  The integrations in the $v$
direction are performed using a three-point Simpson method.  The new
feature of the current scheme relative to the classical one is the
necessity of  calculating of  partial  derivative  $d_v$ in
equation $E2$.  This derivative is calculated using a Savitzky-Golay
smoothing filter.

\section{Results and Discussion}
\label{results}
\subsection{The Evaporation}
\label{results_a}
\begin{figure}[t!]
\centering
\noindent
$(a)$\includegraphics[width=14cm]{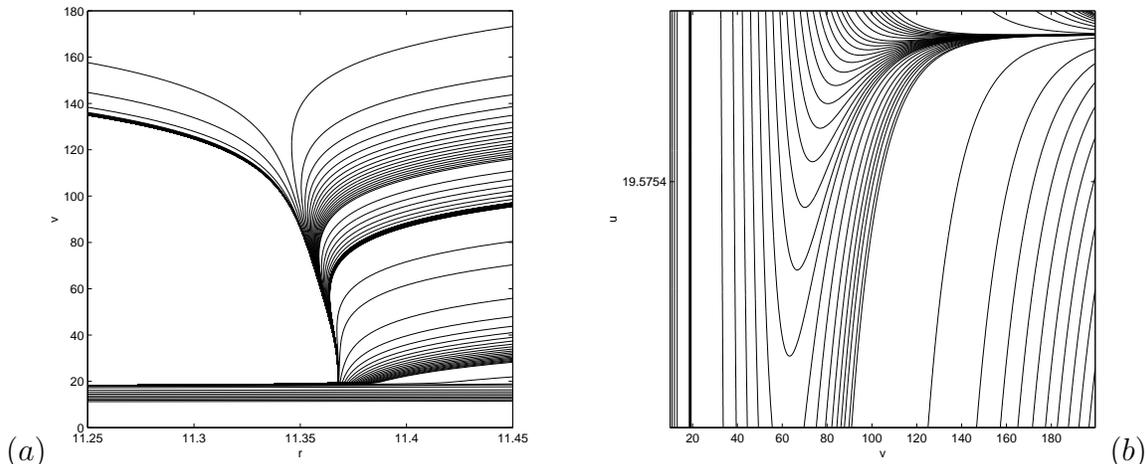} $(b)$
\caption[]{\label{UnchargedEvap}
  
  The evaporation of a Schwarzschild black hole. (a) The 
   radius as a function of $v$ along a sequence of an outgoing
  $u=const$ null rays.  The rays are emitted from the bottom left corner
  at different $u=const$ moments. In the late $u$ times the rays begin
  to curve towards the $r=0$ singularity. There are, however, rays
  which curve back toward the singularity, but then managed to escape
  to infinity, indicating the decrease of mass -- the evaporation. (b)
  The contour lines of the radius in $uv$-plane, in a vicinity of the
  event horizon (defined as a last ray to escape the $r=0$
  singularity).  The radius decreases from the bottom to the top. The
  apparent horizon occurs in the locus of $r_v=0$ and its radius decreases --
  a clear evidence for evaporation.

}
\end{figure}

As a test we check that our code reproduces the uncharged black hole
evaporation \cite{ParPir,AyPir}. To do this, we solve numerically the
set of the coupled Klein-Gordon-Einstein equations.  The complex scalar
field is replaced by a real one.  We set $B=0, e=0$, but $A \ne 0$.  We set
the free parameters to be: $r_0=v_1=10,v_f=200,u_i=0,v_3=20, P=0.01$
and $A=0.65$.  In Figure \ref{UnchargedEvap} we show the dynamical
spacetime obtained in this case.

Figure \ref{UnchargedEvap}(a) depicts the radius as a function of the
outgoing null coordinate $v$ along a sequence of an outgoing $u=const$
null rays. The evaporation is mandatory from this picture. Rays which initially
incline towards the central $r=0$ singularity manage at some later
moment (at the apparent horizon, where $r_v=0$) to escape to infinity.
The evaporation is indicated by the decreasing  radius of the
apparent horizon. Figure \ref{UnchargedEvap}(b) displays the radius
contour lines in the $uv$-plane. The apparent horizon is indicated by
vanishing of the $r_v$ derivative. Again, the decrease of this radius
is clear. In passing, we note that our null coordinates are not very
convenient to resolve the evaporation of a neutral black holes. The
evaporation is contained within a tiny $u$ lapse. The convenient
choice is the null coordinates defined at infinity \cite{ParPir}. In
these coordinates the evaporation takes an infinite (retarded) time.
We did not used these coordinates since our target is the entire
spacetime of an evaporating charged black hole. Therefore, we need a
set of coordinates which are regular on the outer horizon in order to
push the integration inside the black hole.  Our original coordinates,
being Kruskal-like, supply this set. 
We find out that our results are in agreement with the previously
established ones \cite{ParPir,AyPir} and turn to the main concern of
this work -- the evaporation of charged black holes.

\begin{figure}[t!]
\centering
\noindent
\includegraphics[width=14cm]{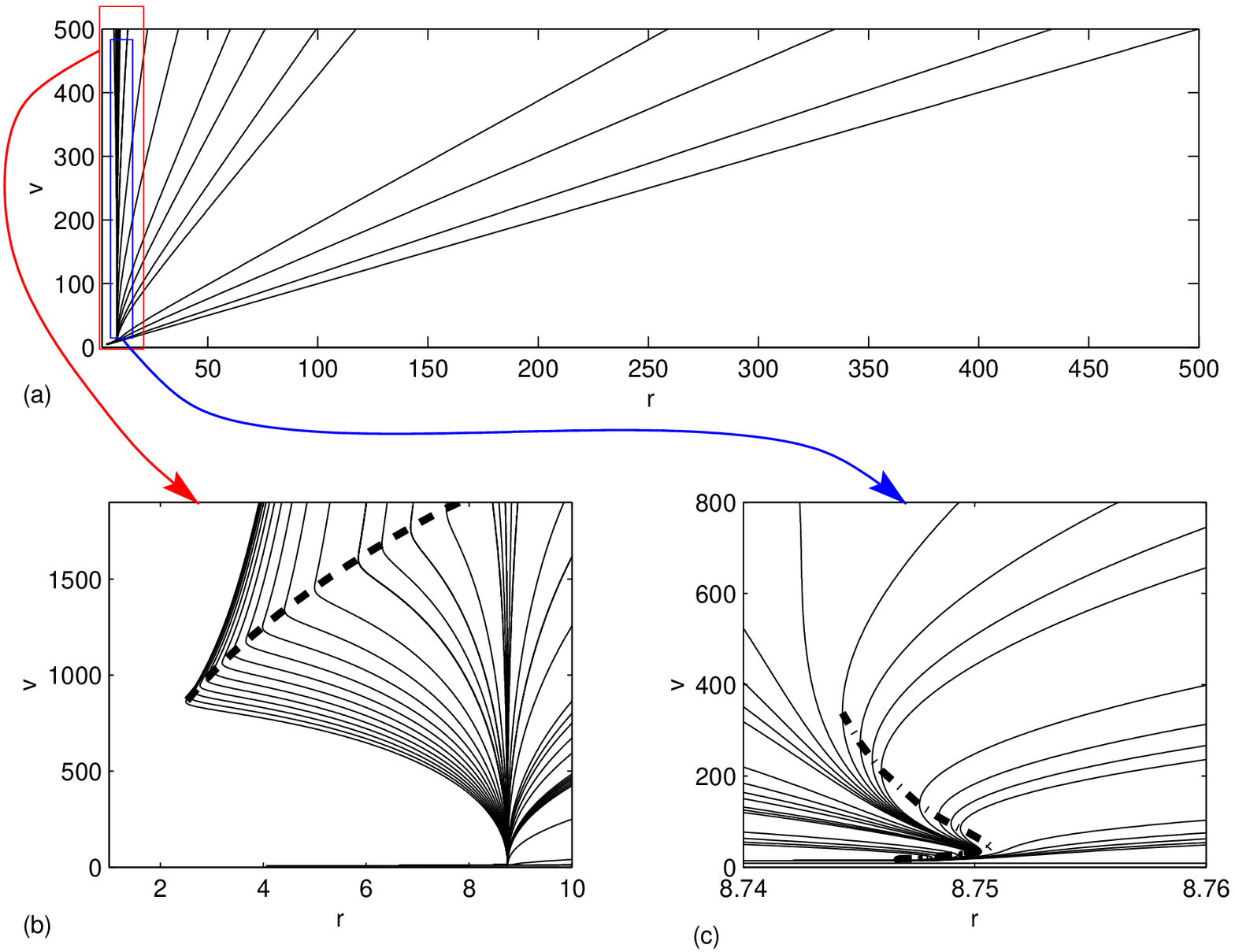}
\caption[]{\label{fig_rad12}
  The evaporating charged black hole. Figures depict radius $r(u,v)$
  as a function of the advanced time $v$ along a sequence of $u=const$
  outgoing null rays. The retarded time $u$ grows from the rightmost
  ray to the leftmost one.  (a) The entire spacetime for a charged
  collapse.  (c) The part of (a) in the vicinity of the outer apparent
  horizon (signaled by vanishing of the $r_v$ derivative), which
  displayed as a thick dashed line. The horizon initially grows,
  indicating absorption of a matter and formation of a black hole.
  After that the horizon contracts, indicating the loss of mass by the
  evaporation. (b) The part of (a) in the vicinity of the inner
  apparent horizon (again it is signaled by vanishing of the $r_v$
  derivative), this horizon is  displayed as a thick dash-dotted line.
  Note that the horizons ``move'' to approach asymptotically ($v
  \rightarrow \infty$) each other.

}
\end{figure}

\begin{figure}
\centering
\noindent
\includegraphics[width=8cm]{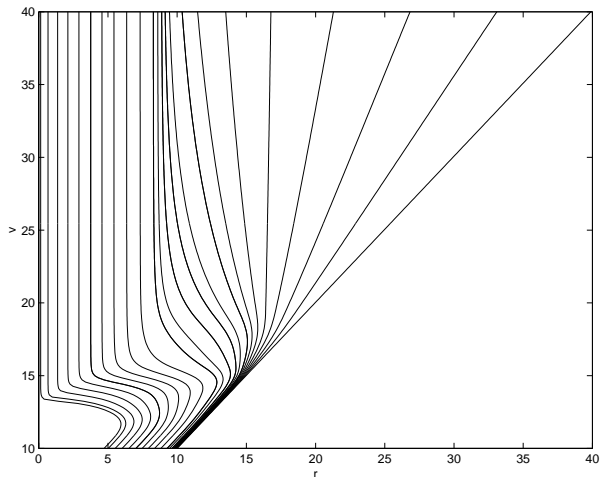}
\caption[]{\label{fig_Classics}
  The classical collapse. The radius $r(u,v)$ as a function of the
  advanced time $v$ along a sequence of $u=const$ outgoing null rays.
  The retarded time $u$ grows from the rightmost ray to the leftmost
  one. There is an outer apparent horizon  located at about $r\approx
  16$ (not shown) and the  CH indicated by the vertical rays (see further
  discussion in Ref. \cite{HodPir,SorPir}). }
\end{figure}

We have set the rest of the free parameters as: $v_1=5, v_2=14, v_2'=v_3,
v_f=2000, A=0.35, B=0.43$ and $e=0.151$. The Figures below are
obtained with this specific choice of the parameters.  Figure
\ref{fig_rad12}(a) displays the radius $r(u,v)$ as a function of the
advanced time $v$ along a sequence of $u=const$ outgoing null rays.
The retarded time $u$ increases from the rightmost ray to the leftmost
one.  The rays are emitted from the surface of a collapsing matter (at
the left bottom corner). The first rays, which escapes to infinity, are
unaffected by the matter indicating the asymptotic flatness of the
spacetime.  As the collapse proceeds the rays begin to curve back
towards the origin.  Without evaporation there would be a ray which
does not escape to infinity, but stays at a constant radius, this ray
indicates the event horizon. The rays emitted after that would
asymptotically ($v \rightarrow\infty$) become vertical indicating the
formation of the Cauchy horizon (CH) \cite{HodPir,SorPir}. Finally,
the rays emitted in late retarded time would inevitably fall into the
strong central singularity. In Figure \ref{fig_Classics} we depict the
${rv}$-diagram obtained in the classical case, a case without
evaporation.

Inclusion of the evaporation changes this picture, specifically the
inner and the outer horizons are not stationary null hypersurfaces. 
Figure \ref{fig_rad12}(c)  depicts the enlarged part of Figure
\ref{fig_rad12}(a) -- the part where the outgoing rays curve to be
swallowed by the formed black hole, but then managed to escape to
infinity. The locus of the turn-points of these outgoing rays
indicates the outer apparent horizon.  Figure \ref{fig_rad12}(b)
displays another enlarged part of Figure \ref{fig_rad12}(a)-- the part
where the outgoing rays tend towards the CH.  The  incline of
these, classically vertical, rays towards the outer horizon (above the 
thick dashed line, see Figure) can be observed. The turn-points
(i.e. vanishing of the $r_v$ derivative) of these rays indicate the
inner apparent horizon.

\begin{figure}[t!]
\centering
\noindent
\includegraphics[width=14cm,height=8cm]{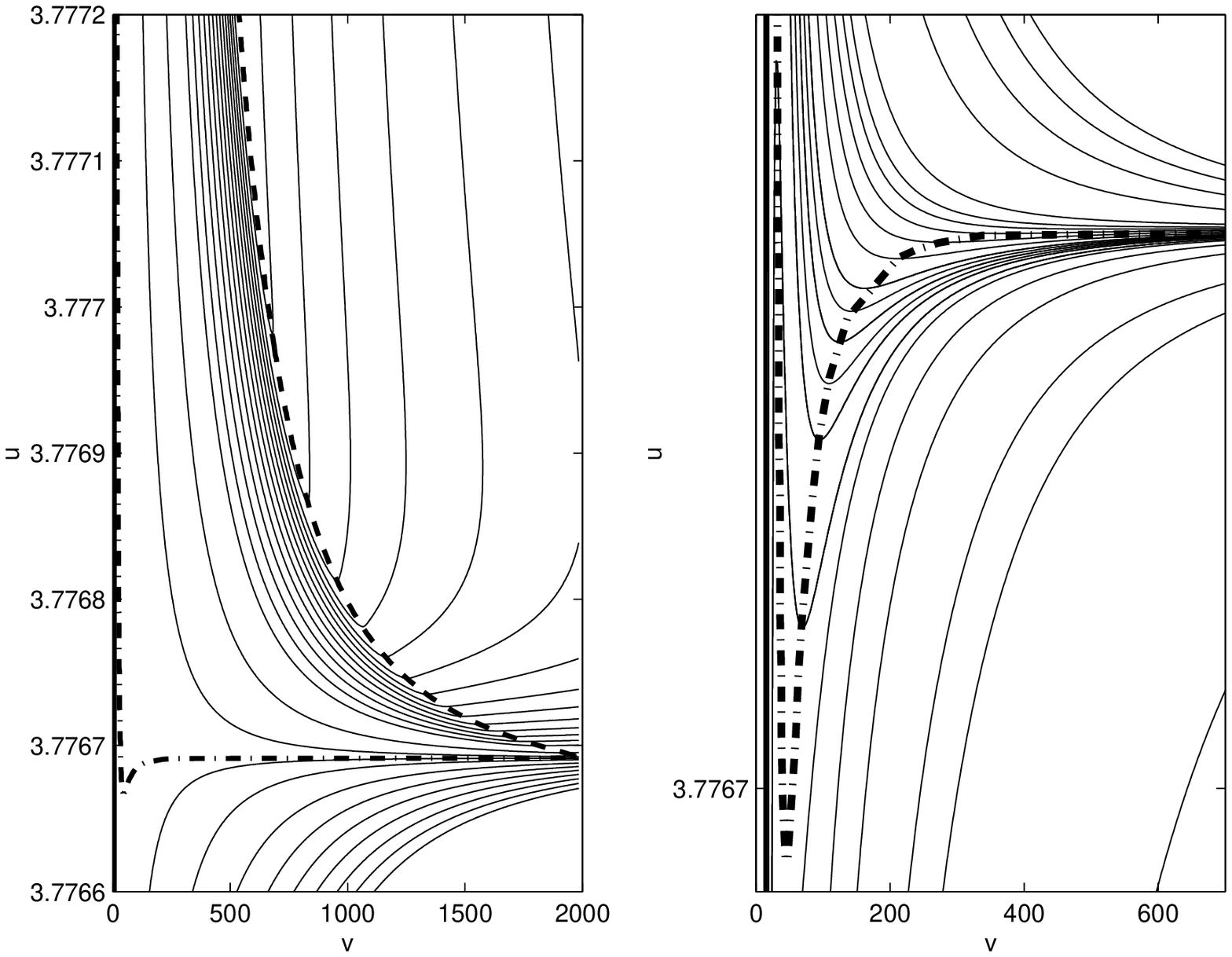}
\caption[]{\label{fig_contour}
  Left panel: lines of constant radius in $uv$-plane. The inner and
  the outer apparent horizons of an evaporating black hole are defined
  by vanishing of the $r_v$ derivative. The inner horizon (shown by
  thick dashed line) is approaching the outer horizon (the thick
  dash-dot line) in the asymptotically late advanced time, $v
  \rightarrow \infty$. Right panel: lines of constant radius in
  $uv$-plane in a vicinity of the event horizon. The outer apparent
  horizon is designated by the thick dash-dot line. This horizon
  contracts from about $v\approx 50$, indicating the mass-evaporation.
  On the both panels the radius is decreasing from the bottom to the
  top.

}
\end{figure}

To gain a further insight into the emerging spacetime we depict in
Figure \ref{fig_contour} the radius contour lines in  $uv$-plane.
We show explicitly  the inner apparent horizon (a thick dashed line)
and the outer apparent horizon (a thick dot-dashed line). The outer
horizon initially grows from zero, that corresponds to the formation
of a black hole and then it shrinks signaling on the
mass-evaporation. The inner horizon (CH) does not remain a contracting 
null hypersurface, as in a case without evaporation, but instead 
it expands outwards to meet asymptotically the outer horizon.
  
The spacetime of the EBH has a very different structure from the
non-extremal one, in particular the outer and the inner horizons are
located at the same radial coordinate. In the evolutionary context,
where the initially non-extremal black hole evolves towards the EBH,
the merge of the both horizons will occur in an infinite advanced time
(as $v \rightarrow \infty$).  The numerical simulation can proceed
only to finite values of $v$, and the horizons will only tend to approach
each other. We indeed observe this during the simulation. The outer
horizon (indicated by tick dotted line in Fig.\ref{fig_rad12}(b))
approaches the inner horizon (indicated by tick dashed line in
Fig.\ref{fig_rad12}(c)). The same picture is seen in Fig.
\ref{fig_contour}.  It should be noted that the shrinking of the outer
apparent horizon is contained within a small variation of the $u$
coordinates, as in a neutral collapse. That's because of a
Kruskal-like nature of our coordinates.

\begin{figure}[h!]
\centering
\noindent
\includegraphics[width=8cm]{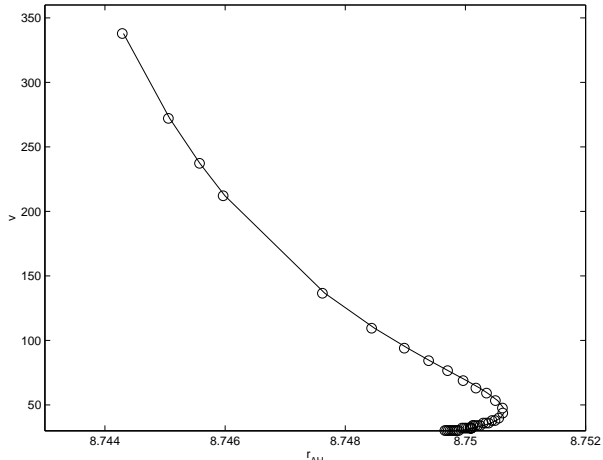}
\caption[]{\label{fig_OuterHorizon}
  The location of the outer apparent horizon of the evaporating black
  hole as a function of the advanced time $v$.The horizon initially
  grows indicating the absorption of matter by the black hole.
  Subsequently (starting from $v\approx 50$) the horizon contracts,
  signaling on evaporation. The circles represent the numerically
  obtained radius of the horizon.  The solid line represents the
  location of the horizon, which was calculated assuming RN metric,
  according to Eqn. (\ref{r_ah_RN}).  The notable similarity of the
  results shows that a charged black hole is indeed a RN black hole
  during its evolution, from a point of view of an observer at
  infinity.
 
}
\end{figure}

It was interesting to compare the shrinking of the radius of the outer
apparent horizon with the radius of this horizon, as it is calculated
assuming the external RN metric. In the latter case the radius of the
horizon is given by:
\begin{equation}
\label{r_ah_RN}
r_{+}=M+\sqrt{M^2-Q^2} \ .
\end{equation}
Here and below $M\equiv M_{\rm AH}(u,v)$ and $Q\equiv Q_{\rm AH}$
is the mass and the charge of the apparent black hole. In Figure
\ref{fig_OuterHorizon} we depict the radius of the outer apparent
horizon vs. the advanced time. One can observe that the numerically
calculated radius of the horizon (designated by the circles in
Figure) and the one obtained by the above formula (solid line)
practically coincide. It must not be confusing, despite this
remarkable result, the entire spacetime of the dynamically evaporating
black hole is very different from the RN one. The radius of the inner
horizon, defined in a RN spacetime as $r_{-}=M-\sqrt{M^2-Q^2} $ does
not coincide with the calculated inner horizon. This is because of
presence of energy influx which changes the mass-function inside the
black hole. This mass-function is different from the external one, $
M_{\rm AH}$.

\begin{figure}[t!]
\centering
\noindent
\includegraphics[width=14cm,height=6cm]{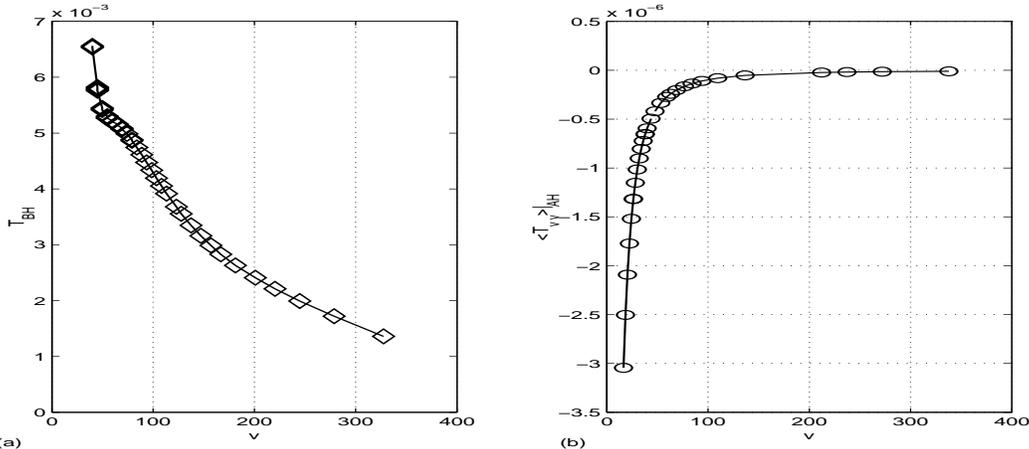}
\caption[]{\label{fig_TTvv}
  (a)The Hawking-Bekenstein temperature $T$ of a black hole as a
  function of an advanced time $v$ along the outer apparent horizon.
  The temperature decreases as the black hole evaporates and
  approaches the EBH state.  (b)The negative influx ${\langle T_{vv}
    \rangle |}_{AH}$ as a function of $v$. The flux tends to the zero
  value in the late advanced time.}
\end{figure}

The Hawking-Bekenstein temperature of a RN black hole, as it measured in
infinity is defined as:
\begin{equation}
\label{Temper}
T={\sqrt{M^2-Q^2}\over {2 \pi r_+}} \ . 
\end{equation}
In Figure \ref{fig_TTvv}(a) we plot the dependence of the  temperature 
of the dynamical black hole on the advanced time $v$.
The temperature is seen to decrease. We note that the obtained black
hole is close to a EBH as $1-Q/M \approx 1/400$. 
 
Another quantity of interest is the  mass-loss rate. This can be
calculated from a negative infalling energy flux $\langle T_{vv}
\rangle$, which expected to drive black hole mass-evaporation
according to:
\begin{equation}
\label{Tvv}
{dM\over dv} = {\langle T_{vv} \rangle |}_{AH}  \ . 
\end{equation}
Figure \ref{fig_TTvv}(b) depicts the explicitly calculated infalling
energy flux $\langle T_{vv} \rangle$ along the outer apparent horizon
as a function of an advanced time $v$.  The negative influx approaches 
zero value with the lapse of the advanced time $v$. This corresponds
to the stable point in the evolution of the black hole.  The black
hole tends towards the EBH state, which would be reached at an infinite
lapse of $v$.

\subsection{ Evaporation with  Discharge}
\label{results_b} 
In the next stage of the analysis we relax the assumption of a stable
charge, allowing it to be radiated away via the Schwinger pair
creation process. We assume that charged particles are not produced in
the Hawking process. We examine analytically the fate of a
simultaneously evaporating and discharging black hole.

We suppose that the discharge begins only when the electric field of
the collapsing matter approaches the critical value, $ E_c$,
and there is no pair creation for the subcritical fields. This is
justified since the rate of pair creation for subcritical fields is
exponentially suppressed (see e.g.\cite{GMM}).

\begin{figure}[t!]
\centering
\noindent
\includegraphics[height=8cm]{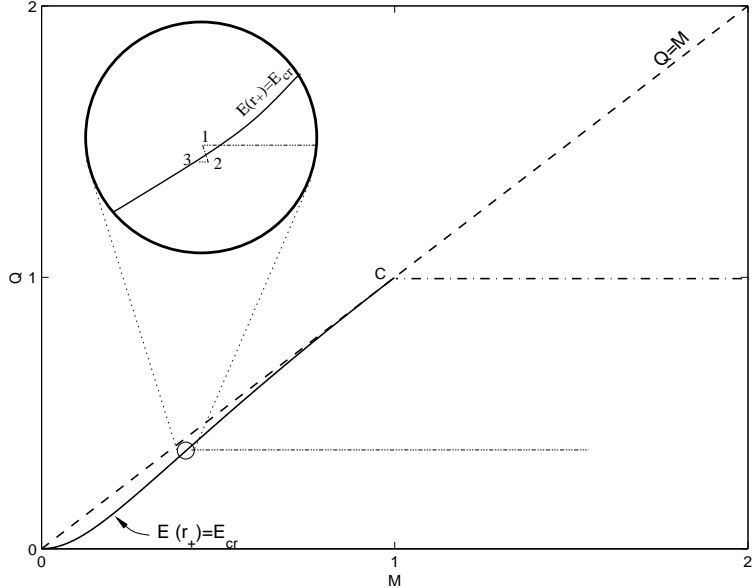}
\caption[]{\label{fig:Q-M}
  A QM-diagram.  Black holes can exist only in region below  $Q=M$ track,
  designated here by the dotted line.  The solid line in the diagram
  corresponds to the black holes which experience an extremal
  pair-creating electric field along their outer horizon.  An
  evaporating black hole is  presented in this diagram as a
  horizontal line. There is the limiting value of charge for an
  evaporating black hole, such that a black hole having charge above
  this value will not discharge but rather settle down to an EBH state.
  We depict a trajectory (dash-dot track) for  a limiting black hole.}
\end{figure}

Figure \ref{fig:Q-M} is the QM-diagram. The black hole can only form
if $Q \le M$ thus only the region below $M=Q$ is allowed.  $M=Q$ is
designated by the dashed line in this Figure. The solid line
represents black holes for which: $E={Q / r_{+}^2}=E_c$ along the
outer horizon. Below this line the electric field upon the outer
horizon is subcritical and charged particles are not produced. Using
(\ref{r_ah_RN}) we find the equation of this trajectory:
\begin{equation}
\label{M(Q)}
M(Q,E_c) = {1 \over 2} \sqrt{Q \over E_c} + {1 \over 2}
\sqrt{ E_c \over Q} Q^2  . 
\end{equation}
 This line   intersects the $Q=M$ EBH line at $Q= E_c^{-1}$. 
 After the intersection the line continues to the unphysical, $Q>M$,  region.
 
 A trajectory of a black hole that evaporates, starting from an
 initial mass and charge, is represented by a horizontal line in this
  diagram. If this trajectory intersects the $E= E_c$
 line it signals  the onset of discharge.  We depict
 such a trajectory (dash-dot line in the diagram) that all the three
 lines intersect at the same point, $Q= E_c^{-1}$, the point C
 in Figure \ref{fig:Q-M}. The black holes that evolve along 
 horizontal trajectories lying above the dash-dot line  never
 experience a critical pair-creating field along their horizon.
 These black holes do not discharge but rather settle down to the
 stable EBH state. The fate of the black holes that
 do experience critical field is completely different.  
 
 We show below that the following scenario holds. Black holes which
 evaporate along trajectories, lying below the dash-dot track reach
 the pair creating field at some moment. Then the black hole evolve
 along the $E=E_c$ line discharging and evaporating simultaneously,
 decaying to a zero size.  If the black hole evaporates exactly on the
 dash-dot track then it moves towards point C. Here we must recall
 that a small number of pairs will be produced even for a subcritical
 field.  This small amount of pairs is sufficient to reduce (just a
 little) the charge of a black hole, pushing it away from the point C
 towards the decaying regime. This behavior will hold also for a small
 region above C as long as the pair creation is not suppressed
 exponentially.

Let us consider an evaporating black hole, moving along a horizontal
line in the region below the dash-dot track. At some moment this black
hole will approach the $E(r_+)=E_c\equiv{Q / r_c^2}$ line. Suppose
that the black have evolved slightly into the region above that line.
The electric field along its outer horizon is now supercritical and
pairs are intensively created. The created pairs reduce the charge of
a black hole to the value $Q_c$ defined by ${ Q_c / r_+^2}=E_c$.  The
pairs being created in the vicinity ($1-{r_+ \over r_c} \ll 1$) of the
outer horizon of the black hole are swallowed by it. This forces the
increase, $\Delta M$, of the mass of a black hole and therefore
increase of the radius of the outer horizon $r_+ \rightarrow
r_+'>r_+$.  The electric field along $r_+$ is exactly critical and
therefore along $r_+'$ is subcritical.  As a result the black hole
moves outwards away from the supercritical region.  The black hole has evolved
$(M,Q) \rightarrow (M' , Q_c)$ with $M'= M+\Delta M$ during the pair
creating stage. In the next stage the black hole evaporates until
reaching (and maybe exceeding) again the critical field along its
horizon. At the end of this phase the black hole would evolve $(M' ,
Q_c) \rightarrow (M'', Q_c)$. If the electric field along the outer
horizon is supercritical the above story takes place once again and so
on until the full decay. The cycle $(M,Q) \rightarrow ( M' , Q_c)
\rightarrow (M'', Q_c)$ is indicated by the cycle 1-2-3 in the
zoom-insert in Figure \ref{fig:Q-M}.  The processes in nature are not
discrete as we presented it but rather continuous and both phases take
place simultaneously.

The conclusion from the above considerations is clear -- the black
holes will evolve along the $E(r_+)=E_c$ track after approaching it. Along
this track it satisfies
\begin{equation}
\label{dMdQ}
 {dM \over dQ} ={1 \over 4} \sqrt{1 \over E_c
  Q}+{3 \over4} \sqrt{ E_c Q} \ ,
\end{equation}
as  obtained by differentiating Eq. (\ref{M(Q)}). 

It is also of interest to calculate the mass- and the charge-loss
rates of a black hole evolving along the $E(r_+)=E_c$ line.  To
address this question we implement the idea that if the electric field
of a black hole exceeds the critical pair-creating value then the
number of created pairs is not simply $N_0={Q_c -Q \over e}$, but
rather enormously amplified. This observation was previously made in
\cite{ruffini}, tough the calculation we present below differs from
that in Ref. \cite{ruffini}.  The energy of the created pairs,
$\epsilon_p$, is extracted from the energy of the electric field of
the black hole.  If the electric field exceeds only slightly the
critical value (that holds for $1-{r_+ \over r_c} \ll 1$ in our
circumstance) then the pairs are created nearly at their rest energy
with very little kinetic energy. This is in fact the definition of the
critical pair-creating field. Therefore, the number of the created
pairs can be estimated as $ N_p\approx{\epsilon_p / m}=\Delta M /m$ and
$N_p \gg N_0$ (holding for $e/m \gg 1$). The pairs are created in the
region $(r_+,r_c)$ with the proper thickness of $\delta r \equiv
r_c-r_+$.  The pairs constituting that region are in a state of the
quasi-neutral plasma.  Only the outermost and the innermost layers of
this shell have an uncompensated charge of $Q-Q_c$ and the number of
pairs in these layers is indeed $N_0$ \cite{ruffini}.

The rate of the pair creating process at a given radius within the
pair-creating region can be approximated as
\begin{equation}
\label{rate}
 \Gamma={d N \over \sqrt{-g} d^4x}= {1\over 4 \pi} {\left(e E \over
   \pi \right)}^2 e^{-{\pi E_c \over E}} \ .
\end{equation}
From this we can estimate the time
needed to create  the $N_p$ pairs as
 \begin{equation}
\label{dT}
\Delta\tau_{dis} \approx{ N_p\over \Gamma V} \ ,
\end{equation} 
where $V$ is the proper volume of the pair creating region.  Using the
above results we can estimate the mass increase and the charge loss by
the black hole due to discharge. 

The charge loss is estimated as $\dot Q = e {dN_0\over
  d\tau}\approx{Q-Q_c \over \Delta\tau_{dis}}\approx{Q-Q_c \over
  N_p}\Gamma V $.  The total energy of pairs and therefore the mass
increase, $\Delta M$, of the black hole is estimated knowing the
initial, $Q/r^2$, and the final, $Q_c/r^2$ electric field
configurations
\begin{equation}
\label{dM}
 \Delta M  \approx {Q^2 -Q_c^2\over 2 r_+} \ .
\end{equation}
Here we have ignored the energy carried by the $N_0$ particles,
repulsed to infinity since $N_p \gg N_0$. Using Eq. (\ref{dM}) to
calculate $N_p =\Delta M / m$ one obtains, using $\delta r \ll r_+$,
the charge loss
  \begin{equation}
\label{Qdot}
\dot Q \approx {2 r_+ \over Q+Q_c} m \Gamma V  \approx { r_+ \over Q} m\Gamma V   \ .
\end{equation}

In  the evaporation phase  the mass is radiated
away by the Hawking process. In order to estimate this rate of mass
loss we assume that the black hole radiates as a black body with
temperature T, defined by (\ref{Temper}), and surface area ${ \cal A} = 4
\pi r_+^2$. Hence we obtain the approximate mass-evaporation rate:
\begin{equation}
\label{Hrate}
{\dot M}_{ev} \equiv H \approx \sigma T^4  \cal A   \ ,
\end{equation}
where $\sigma$ is the Stefan-Boltzmann constant.

Both processes of evaporation and discharge are simultaneous. The
evaporating black hole approaches and exceeds slightly, say by $\delta
r$, the critical radius $r_c$. The value of $\delta r$, which defines
the volume of the pair-creating region according to $V=4 \pi
r_+^2 \delta r$, will be determined by the requirement that the ratio
${\dot M} / {\dot Q}$ is given by (\ref{dMdQ}). That is  the black hole
evolves nearly along the $E(r_+)=E_c$ line.  The rate of the mass loss
is given by $ \dot{M} =\dot{M}_{ev}-\dot{M}_{dis} = H - {\Delta M
  \over \Delta\tau_{dis}}\approx H- m \Gamma V$. The rate of the charge loss
is given by Eq. (\ref{Qdot}).  Therefore, from
\begin{equation}
\label{rate1}
\left({dM \over dQ}\right) \equiv {\dot{M} \over \dot{Q}}={H- m
  \Gamma V \over  {\left( r_+ \over Q \right)} m \Gamma V} \ ,
\end{equation}
assuming that  $\left({dM \over dQ}\right)={\left({dM \over
      dQ}\right)}_{ E(r_+)=E_c}$ one obtains
\begin{equation}
\label{rate2}
\dot{M}=H\left(1-{1\over {r_+ \over Q} {\left({dM \over dQ}\right)}_{ E(r_+)=E_c}+1}\right) \ .
\end{equation}
With the use of Eqs.(\ref{M(Q)}), (\ref{dMdQ}) and (\ref{Hrate}) this
determines the rate of the mass loss in the combined
discharge-evaporation process. We have also to perform the necessary
consistency check which assures that $\delta r /r_+ \ll 1$. From Eqs.
(\ref{rate1}),(\ref{rate2}) and (\ref{rate}) with
$\Gamma(E)\approx\Gamma(E_c)$ one finds that $\delta r /r_+ \ll 1$ if
$1/M m \ll 1$ i.e. if $\lambdabar / r_+ \ll 1$, where $\lambdabar$ is the
Compton wavelength of the created particles. The last inequality holds
until the very late stages of the decay when the discharge occurs also
via the Hawking process. Hence  until the late phases of decay our
calculation remains reliable.

Since ${r_+ \over Q}\left({dM \over dQ}\right)_{ E(r_+)=E_c} $ is
positive and both ${r_+ \over Q}>1$ and $\left({dM \over dQ}\right)_{
  E(r_+)=E_c} >1 $,  the rate (\ref{rate2}) of decay of
charged black hole is slower than that of the uncharged evaporating
one $\dot{ M} <H  $ and together $H>\dot{ M }>H/2$.  Therefore the life-time of the evaporating and discharging black
hole is {\it longer} than the life-time of an evaporating neutral
black hole of the same mass.  This is readily understood -- what we
have after all is the conversion of electrostatic energy of a black
hole to the mass, which is radiated away as a Hawking radiation.
Since that effective mass exceeds $M$ (approximately by the energy of
the electric field), the process of decay lasts more time.

Clearly the decay as we describe it is valid only above the Planck
scales, where our (semiclassical) approximation breaks down in favor
of full Quantum gravity.

\section{Summary and Conclusion}
\label{conclusions}
We have investigated the formation and evaporation of charged black
holes.  In the first stage of analysis we assumed stability of the
charge, while allowing an evaporation and a decrease of the mass via
Hawking radiation. The evaporation was modeled by introducing the
expectation value of a 2D stress-energy tensor of a 2D quantized
scalar field as a source to the semiclassical Einstein equations.  By
doing so, we disregard possible contributions which arise from the
dimensional reduction procedure \cite{BalFa}. The source which we
utilized is (at least) a part of the full 4D stress-energy tensor, and
it gives a feeling of the full back-reaction.

We find that the evaporating black hole loses its mass and approaches
asymptotically a stable endpoint corresponding to the extremal black
hole. Starting from an initially regular matter distribution we have
dynamically obtained a non extremal black hole and than, after
evaporation, a near extremal one.  The outcome of our analysis is the
entire\footnote{Actually, our solution does not include the origin of
  coordinates, $r=0$, and the asymptotically late advanced time
  region. The first is not included in a sake of simplicity of the
  numerics, while the second is not included since it has an
  infinite extension. In any case the region which
  is obtained  supply enough information to draw the conclusions, which
  we have drawn.  } dynamical spacetime of such a black hole.

The spacetime structure of a non extremal and an extremal black holes
are very different. The most striking feature of the latter is that
its inner and outer apparent horizons are located at the same radial
coordinate. In the dynamical context we observe that the horizons tend
to approach each other in a late advanced time. The actual merge will
occur indeed in an {\it infinite} (advanced) time, i.e. as
$v\rightarrow\infty$. Thisis in agreement with  the unattainability
formulation of the third law of the black hole thermodynamics. The
stability of the final EBH state is manifest from the vanishing of the
negative energy influx, $\langle T_{vv}\rangle$, governing the black
hole evaporation according to Eq.  (\ref{Tvv}). The Hawking-Bekenstein
temperature decreases as the black hole approaches the EBH state.

We have also considered a collapse with evaporation accompanied by a
discharge due to instability of the vacuum in strong electric fields.
In this case there is a limiting value of charge for a given mass such
that black holes with a charge above this value would not discharge
but rather evaporate towards a stable EBH state.  Black holes with a
charge below the limiting charge evaporate and discharge
simultaneously, decaying to zero size. It is interesting to note that
such black holes are pushed away from the EBH state. The discharge
thus prevents from small black holes to become extremal. The life time
of an evaporating and discharging black hole is longer then that of an
uncharged black hole of the same mass that decays to zero emitting
Hawking radiation.

Note added: When this work was completed its brought to our attention
that there are related works \cite{new} considering evaporation of 2D
RN black holes.
 
\label{discussion} 

\noindent
{\bf ACKNOWLEDGMENTS:} The research was supported by a grant from the Israel
Basic Research Foundation.

\vspace{0.5cm}
\noindent $^{\ast}$Electronic address: sorkin@merger.phys.huji.ac.il

\noindent $^{\dagger}$Electronic address: tsvi@nikki.phys.huji.ac.il


\begin{thebibliography}{99}

\bibitem{isr}
W. Israel, Phys.Rev.Lett. {\bf 57}, 397 (1986).

\bibitem{ted}
T. Jacobson,  Phys.Rev. {\bf D57}, 4890 (1998). 

\bibitem{StTri}
A. Strominger and S.P. Trivedi, Phys.Rev.{\bf D48}, 5778 (1993).

\bibitem{LooL}
D.A. Lowe and M. O'Loughlin, Phys.Rev. {\bf D48}, 3735 (1993). 

\bibitem{BirDev}
N.D. Birrell and P.C.W. Davies,{\it Quantum Fields In Curved Space},
Cambridge University Press 1982.
\bibitem{DFU} P.C.W. Davies, S.A. Fulling, W.G. Unruh, Phys.Rev. {\bf
    D13}, 2720 (1976). 
\bibitem{Page} D. Page, Phys.Rev. {\bf D13}, 198 (1976).

\bibitem{BalFa} For a review see: R. Balbinot
  and A. Fabbri, Phys.Rev.{\bf D59} 044031 (1999).

\bibitem{ParPir} R. Parentani and T. Piran, Phys. Rev. Lett. {\bf 73},
  2805 (1994).



\bibitem{AyPir} S. Ayal and T. Piran, Phys.Rev. {\bf D56}, 4768 (1997).

\bibitem{HamSt} R.S. Hamad\'e and J.M. Stewart, Class. Quantum Grav.
  {\bf 13}, 497 (1996).  

\bibitem{HodPir} S. Hod and T. Piran, Phys.
  Rev. Lett. {\bf 81}, 1554 (1998); Gen. Rel. Grav. {\bf 30}, 1555
  (1998).  

\bibitem{SorPir} E. Sorkin and T. Piran, Phys.Rev.{\bf D63} 084006 (2001).

\bibitem{BurOr} L.M. Burko
  and A. Ori, Phys.Rev.  {\bf D56}, 7820 (1997).  

\bibitem{NumRec}
  W.H. Press, S.A. Teukolsky, W.T. Vetterling and B.P. Flannery, {\it
    Numerical Recipes in Fortran: The Art of Scientific Computing},
  2nd ed. Cambridge University Press 1992.  

\bibitem{GMM} A.A. Grib,
  S.G. Mamaev, V.M. Mostepanenko, {\it Quantum Effects in Strong
    Fields}, Atomizdat, Moscow, 2nd ed. 1988 (in Russian).

\bibitem{ruffini} G. Preparata, R. Ruffini, S.S. Xue,
  astro-ph/9810182; R. Ruffini,
  astro-ph/9811232.

\bibitem{new}
A.Fabbri, D.J.Navarro, J.Navarro-Salas Phys.Rev.Lett. {\bf 85} (2000)
2434 ; and Nucl.Phys. {\bf B595} (2001) 381.
\end{thebibliography}
\end{document}